# Robust Cloud Suppression and Anomaly Detection in Time-lapse Thermography

Christopher Small[1] and Daniel Sousa[2]

[1] Lamont-Doherty Earth Observatory, Columbia University; csmall@columbia.edu
[2] Department of Geography, San Diego State University; dan.sousa@sdsu.edu

*Due to their transient nature, clouds represent anomalies relative to the underlying landscape of interest. Hence, the challenge of cloud identification can be considered a specific case in the more general problem of anomaly detection. The confounding effects of transient anomalies are particularly troublesome for spatiotemporal analysis of land surface processes. Spatiotemporal characterization provides a statistical basis to quantify the most significant temporal patterns and their spatial distributions – without the need for a priori assumptions about the form, amplitude or timing of the observed changes. The objective of this study is to implement and evaluate a robust approach to distinguish clouds and other transient anomalies from diurnal and annual thermal cycles observed with time-lapse thermography. The approach uses Robust Principal Component Analysis (RPCA) to statistically distinguish low rank and sparse components of the land surface temperature image time series, followed by a spatiotemporal characterization of the L component time series to quantify the dominant diurnal and annual thermal cycles in the study area. The RPCA effectively segregated clouds, sensor anomalies, swath gaps, geospatial displacements and transient thermal anomalies into the sparse component time series. Spatiotemporal characterization of the low rank component time series clearly resolves a variety of diurnal and annual thermal cycles for different land covers and water bodies.*

## Introduction

Cloud cover is a persistent scourge for optical and thermal remote sensing. Particularly in the tropics where it is ubiquitous. Opaque clouds represent a complete loss of information on the underlying target of interest, while cloud shadow and partially translucent clouds (e.g. cirrus) can corrupt the recovered radiance signal in ways that are difficult or impossible to correct. The most common approach to cloud contamination is to identify corrupted acquisitions (e.g.[1]) and exclude them from analysis. However, the presence of partial cloud cover within an acquisition does not preclude use of uncorrupted optical and/or thermal signals in areas without cloud cover or shadow. In these cases, cloud masking algorithms often present a partial solution to the problem.

While the objective of cloud masking is generally to exclude corrupted scene elements from analysis, the principal challenge is in robust identification. Both in detecting the presence or absence of a cloud and in determining its spatial extent. While some types of clouds have relatively distinct edges, others grade continuously from opaque to translucent, requiring threshold criteria to determine mask extent [2]. Identification of clouds is further confounded by the wide range of reflectances [3-6] and emissivities [7-9] clouds possess. The problem is therefore to distinguish clouds from background Earth surface on the basis of contrast in reflectance and/or brightness temperature (e.g.[10-12]). The basis of the challenge is the combined variability of the physical properties of both clouds and backgrounds.

Due to their transient nature, clouds represent anomalies relative to the underlying landscape of interest. Hence, the challenge of cloud identification can be considered a specific case in the more general problem of anomaly detection. Whereas clouds are transient anomalies generally excluded from analysis, other types of anomalies may be the focus of an analysis. Indeed, multitemporal remote sensing is often cast as an exercise in anomaly detection when the objective is monitoring change. At the heart of anomaly detection is anomaly definition. Specifically, defining what is the expected background signal and what is a transient anomaly. Regardless of whether the focus is on the background process or the anomalies, the two must be distinguished in a robust manner.

The confounding effects of transient anomalies are particularly troublesome for spatiotemporal analysis of land surface processes because anomalies can obscure the spatiotemporal manifestation of the processes of interest. Spatiotemporal characterization provides a statistical basis to quantify the most significant temporal patterns and their spatial distributions – without the need for *a priori* assumptions about the form, amplitude or timing of the observed changes [13]. In the case of time-lapse thermography, clouds and other anomalies are superimposed on the diurnal and annual thermal cycles resulting from Earth's rotation and orbit. Spatiotemporal characterization might be used to quantify these thermal cycles in the presence of clouds and other anomalies.

The objective of this study is to implement and evaluate a robust approach to distinguish clouds and other transient anomalies from diurnal and annual thermal cycles observed using time-lapse thermography. Specifically, to characterize the thermal cycles of a diversity of land covers and water bodies in the New York metropolitan area between 2018 and 2023 using land surface temperature retrievals from NASA's ECOsystem Spaceborne Thermal Radiometer Experiment on Space Station (ECOSTRESS) sensor. While the methods applied in this study can be directly extended to multitemporal analysis of other air and spaceborne thermal imagery, they could also be applied to ground-based time-lapse thermography where transient anomalies need be distinguished from background thermal cycles (e.g. [14]).

**Data**

Thermal image data were collected by NASA's Ecosystem Spaceborne Thermal Radiometer Experiment on the International Space Station (ECOSTRESS) [15]. The ECOSTRESS instrument is capable of measuring thermal infrared (TIR) radiance in five channels, centered on: 1) 8.29 μm ; 2) 8.78 μm ; 3) 9.20 μm ; 4) 10.49 μm ; and 5) 12.09 μm [16]. ECOSTRESS instrument operates with f/2 optics in a push-whisk configuration, directing light to 8x16x256 single-bandgap Mercury Cadmium Telluride-based focal plane arrays maintained at 60 K. Each ECOSTRESS data granule is comprised of radiance measurements collected by these multiple focal plane arrays, which have been combined to produce a single composite image [17]. Given the International Space Station's orbital altitude of roughly 400 km and the ECOSTRESS swath width of 53°, the spatial extent of each ECOSTRESS granule is approximately 400 x 400 km. Ground sampling distance at nadir is stated to be approximately 38 x 69 m at collection, and 75 x 69 m for derived product. The elliptical pixel footprints are largely a result of high scanning velocity during the 31.6 μs sensor dwell time producing smearing along the scanning direction. Down-sampling by a factor of 2 via pixel binning is applied during ground processing in order to generate pixels with more approximately square dimensions [18]. The precessing nature of the ISS orbit results in highly variable local overpass time, sampling throughout the diurnal heating and cooling cycle.

This analysis uses the ECOSTRESS Level-2 Land Surface Temperature (LST) and Emissivity data product [19]. The conceptual foundation for the derivation of LST data product is rooted in a temperature emissivity separation (TES) hybrid approach[20]. TIR radiances are first produced by correcting at-sensor radiance on a pixel-by-pixel basis using a radiative transfer model. An empirical relationship is then used to predict the minimum emissivity using the minimum-maximum difference (MMD) method [21, 22] LST uncertainty estimates and quality flags are provided with the data.

The New York metro area (NYC) study area spans western Long Island, the five boroughs of New York City and part of northeastern New Jersey. Major water bodies include the Hudson and East rivers, Long Island Sound, New York Harbor and the Atlantic Ocean. The area contains abundant tidal wetlands in Jamaica Bay, the New Jersey Meadowlands, Pelham Bay and multiple wetlands on the periphery of Staten Island. Elevations ranging from sea level to ~80 meters above sea level on the Staten Island highlands and Hudson Palisades allow for a variety of environments and vegetation communities. A vegetation phenology map derived from spatiotemporal analysis of Sentinel 2 vegetation fraction time series (Figure 1a) clearly distinguishes grass, deciduous trees and wetland vegetation communities. The range of elevations and coastal proximities results in a range of phenologies with most deciduous street trees and urban parks greening in mid/late April and some higher elevation forests greening from late May to early June. Wetland vegetation communities green later and senesce earlier than deciduous trees, which retain leaves in various stages of senescence well into November. Grasses remain green year round with increases in both spring and fall. These spatiotemporal variations in vegetation phenology and abundance exert considerable influence on surface energy fluxes on both diurnal and annual time scales.

Spatiotemporal variability of the ECOSTRESS Land Surface Temperature (LST) time series is dominated by the difference between land and water thermal cycles. The greater thermal inertia of the water bodies reduces their variability on both annual and diurnal time scales. The temporal moment composite in Figure 1b illustrates the contrast between higher and lower thermal variability of different land cover types – primarily in the annual cycle which is strongly modulated by the combination of both vegetation phenology and solar illumination geometry and duration.

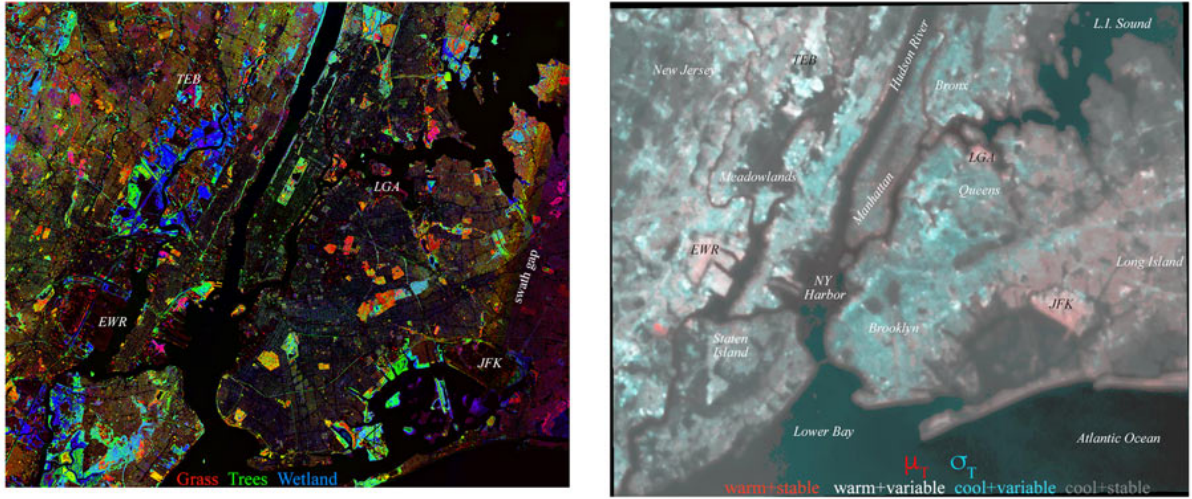

*Figure 1 Index maps for New York metro study area. Vegetation phenology (left) derived from spatiotemporal analysis of Sentinel 2 reflectance shows a continuum of vegetation communities spanning varying mixtures of grass, deciduous trees and wetland vegetation. Fractional vegetation cover modulates both seasonal and diurnal thermal cycles through shading and evapotranspiration. Temporal moment composite of ECOSTRESS Land Surface Temperature (LST) between 2018 and 2023 shows temporal mean LST (mT) and standard deviation (sT), highlighting spatial variations in seasonal and diurnal thermal cycles related to land cover and water body thermal responses. Major airports include Kennedy (JFK) La Guardia (LGA), Newark (EWR) and Teterboro (TEB).*

The orbit of the International Space Station (ISS) provides ECOSTRESS acquisitions at all times of the day and night throughout the year (Figure 2). Between 2018 and 2023, approximately 220 potentially usable (partial cloud cover) acquisitions are available for the NYC study area. Of these, 124 acquisitions could be sufficiently geometrically rectified to allow for spatiotemporal analysis. The LST scenes were acquired from the NASA EarthData portal using the Application for Extracting and Exploring Analysis Ready Samples (AppEARS) tool and were rectified using the Automated and Robust Open-Source Image Co-registration Software (AROSICS) available from: https://github.com/GFZ/arosics. Of these 124 acquisitions, 55 are visually cloud-free and 19 have swath edge gaps – primarily in the NE corner. The dates and times of potentially usable and geospatially rectifiable acquisitions are shown in Figure 2. Three of the rectifiable acquisitions have conspicuous displacements > 1000 m and numerous others have smaller displacements. The combination of the ISS orbit and the distribution of rectifiable acquisitions results in a pronounced day-night asymmetry in the resulting image time series with many more daytime (11-24 UTC) than nighttime (0-10 UTC) acquisitions. This introduces an unavoidable aliasing of the diurnal thermal cycle relative to the more evenly sampled annual cycle. The implications of this are discussed below.

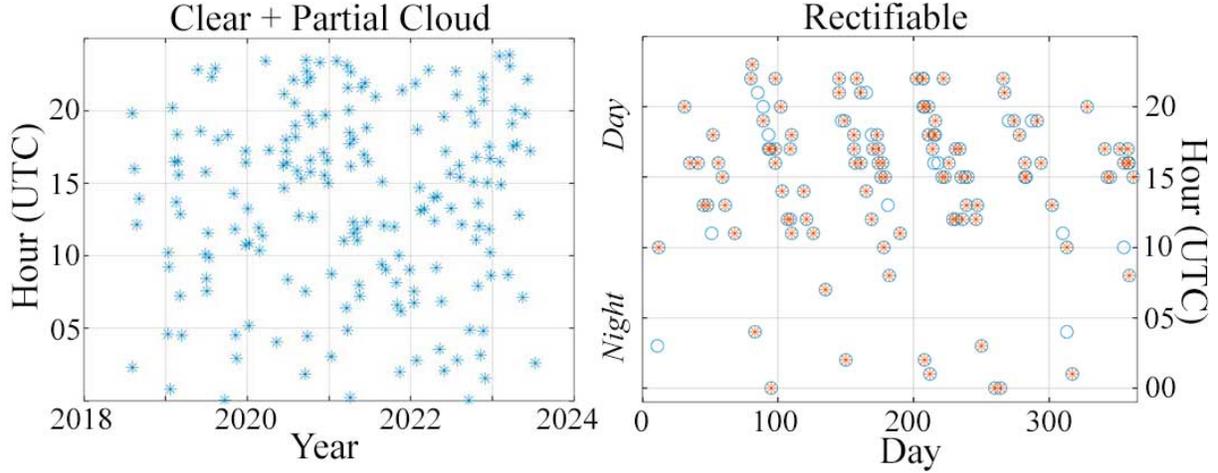

*Figure 2 ECOSTRESS LST temporal coverage for the NYC study area. The ISS orbit provides ~220 potentially usable granules between mid-2018 and mid-2023, spanning all seasons and overpass times (left). Of these, 124 could be sufficiently rectified to allow for multitemporal analysis over a common calendar year (right). Open circles indicate granules with significant swath gaps - primarily in the northeast quadrant. Three of the open circles also indicate unrectified granules displaced ~1000 - 1600 m NE of the other granules. NYC is UTC-5.*

**Methods**

The approach in this study uses Robust Principal Component Analysis (RPCA) to statistically distinguish low rank (L) and sparse (S) components of the LST image time series, followed by a spatiotemporal characterization of the L component time series to quantify the dominant diurnal and annual thermal cycles in the study area. Conceptually, the RPCA separates the low rank structure of the variance associated with the diurnal and annual thermal cycles present in all the LST images from the sparse structure of the less coherent variance associated with clouds, swath gaps and other transient anomalies in a subset of the LST images. As such, it does not require a thermal definition of what constitutes a cloud (as cloud masking approaches do), but rather a spatiotemporal distinction between the underlying thermal cycles of the land surface and the transient anomalies associated with clouds, gaps and other features that are not consistent with the pervasive thermal cycles responsible for majority of the time series' covariance structure.

Robust Principal Component Analysis [23] defines a spatial array of LST time series $M$ to be the sum of a low-rank matrix $L$ and a sparse matrix $S$:

$$M = L + S \qquad (1)$$

The matrix decomposition is generally ill-posed (NP-hard), but under weak assumptions [23] prove that $L$ and $S$ can be recovered exactly through a convex optimization called Principal Component Pursuit. This is achieved by optimizing the sum of the nuclear norm of $L$ ($L_*$, sum of the singular values) and the weighted $L_1$ norm of $S$:

$$\begin{aligned} minimize \quad & \|L\|_* + \lambda \|S\|_1 \\ subject\ to \quad & L + S = M \end{aligned} \qquad (2)$$

where

$$\lambda = \frac{1}{\sqrt{n_{(1)}}}, n_{(1)} = \max(n_1, n_2)$$

(3)

and L is a general rectangular matrix of dimensions $n_1$ x $n_2$. A notable benefit of the RPCA is that no tuning parameters are required.

In this analysis, we implement RPCA using the alternating direction method of [24] as implemented in the R package "rpca" (https://cran.r-project.org/web/packages/rpca/index.html). We use the values of λ and μ (augmented Lagrange multiplier parameter) suggested by [23]. The terminal δ was set to 1 x $10^{-7}$, which reached convergence within 5000 iterations in every case, and often within 1500 iterations, for image time series with a spatial extent of ~ 6.5 x $10^5$ pixels and > 120 temporal dimensions. This corresponded to ≈ 24-48 hour processing times using a quad-core 2 GHz Intel Core i5 CPU with 32 GB of RAM.

The spatiotemporal characterization of the L component time series uses a standard $L^2$ principal component transformation to identify the dominant temporal patterns (annual and diurnal thermal cycles) and their spatial distributions corresponding to different land covers and water bodies. The variance partition of the orthogonal PC dimensions, given by the eigenvalues of the covariance matrix of the time series, provides an indication of the spatiotemporal dimensionality of thermal cycles. This characterization of the spatiotemporal dimensionality and temporal basis functions is equivalent to the Empirical Orthogonal Function (EOF) analysis devised by [25], except that the EOFs are temporal and the corresponding PCs are spatial – as required by much greater number of spatial pixels than temporal acquisition dates. This characterization process is described in detail by [13]. The end result of this characterization is a parsimonious representation of the low rank component of the LST image time series in the form of spatial principal components. These PCs illustrate the dominant thermal cycles on land and water in both geographic space and in a temporal feature space illustrating the diversity of thermal cycles observed. Because of the aforementioned aliasing of the diurnal cycle, these patterns represent primarily the annual cycles.

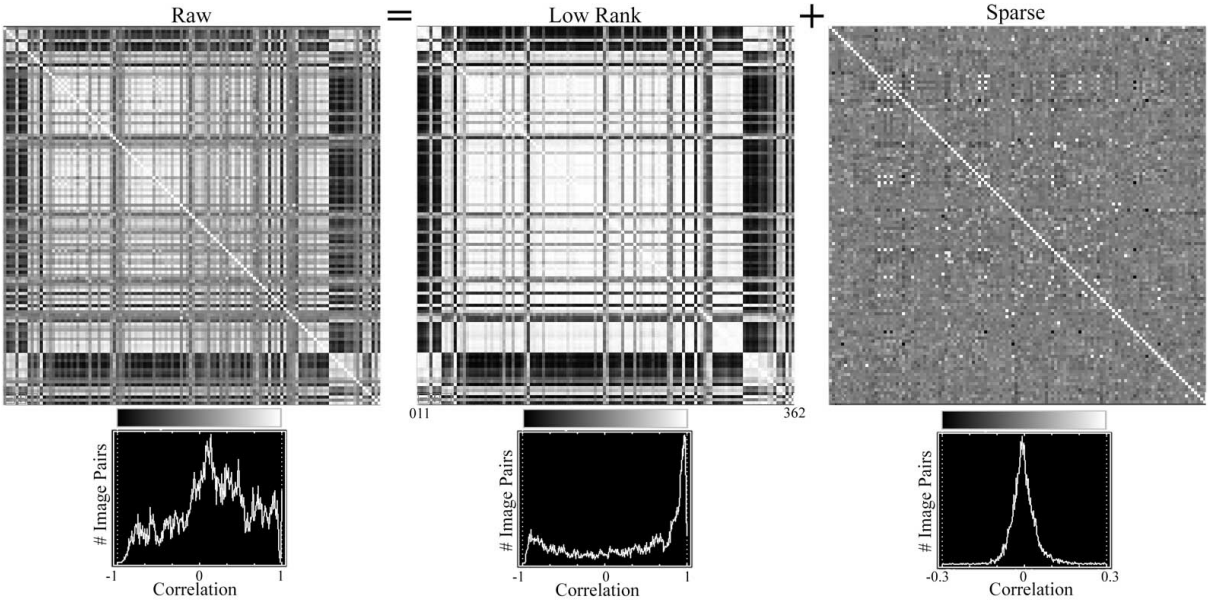

*Figure 3  Correlation matrices and distributions for ECOSTRESS LST time series with low rank and sparse components.  Raw time series contains several acquisitions with swath gaps and spatial displacements of varying distance causing lower spatial correlations.  Higher spatial correlations result from the persistent thermal contrast between land and water.  Both raw and low rank time series have negative correlations for nighttime and winter acquisitions.  The sparse residual time series represents spatiotemporal transients with almost all correlations <  |0.1|.*

**Results**

*Spatiotemporal Characterization*

Spatial correlations among LST acquisitions span the range between -1 and 1, driven by the aforementioned thermal contrast between land and water on both diurnal and annual time scales. Figure 3 shows correlation matrices and distributions for the raw LST time series and the low rank and sparse components.  Anticorrelations near -1 are apparent between late night and winter acquisitions when the land surface is cooler than most water bodies and the majority of daytime acquisitions for which the land surface is warmer.  Low correlations occur in early evening temperate acquisitions in which the water and land surfaces are more nearly isothermal.  The presence of clouds and swath gaps reduces spatial correlations among acquisitions, resulting in a modal correlation near 0 for the raw LST time series.  In contrast, the low rank component time series has a strongly bimodal distribution while the sparse component time series shows the vast majority of correlations within ±0.1 of 0.  The contrast between anticorrelated pre-sunrise winter and mid-afternoon spring acquisitions is illustrated in Figure 4.

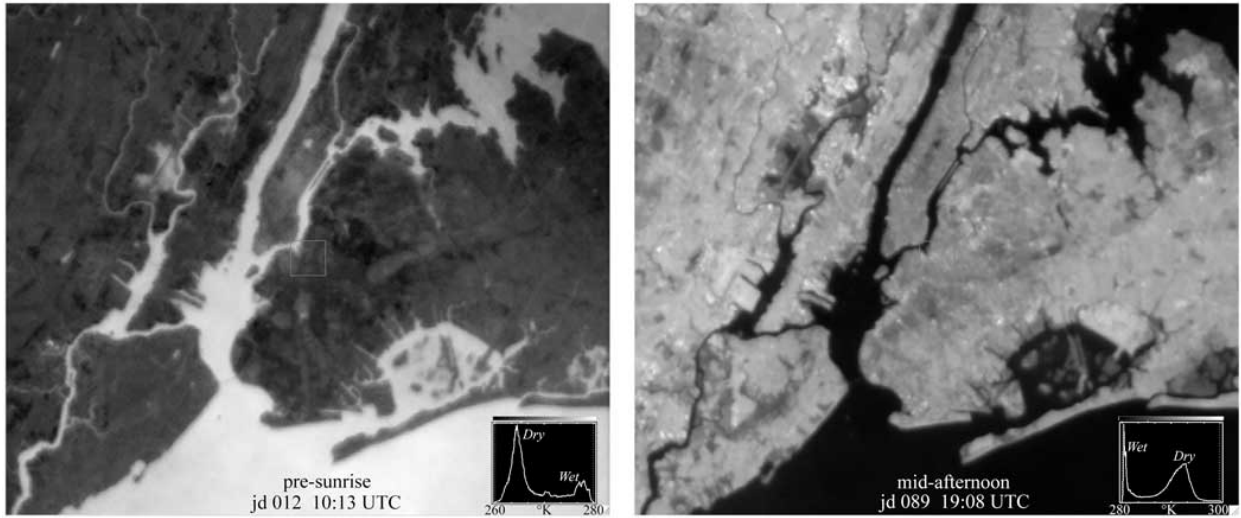

*Figure 4 Seasonal and diurnal thermal contrast. Early morning winter acquisition (left) shows water significantly warmer than land surface. Mid-afternoon spring acquisition (right) shows a bimodal distribution of opposite polarity with a very different spatial temperature distribution on the land surface. Both acquisitions span ~20° K. The spatial correlation is -0.77.*

The variance partition given by the eigenvalues of the principal components (PCs) of the low rank and sparse component time series quantify the effectiveness with which the RPCA separates anomalies and noise from the annual and diurnal thermal cycles. Figure 5 compares this variance partition for the full LST time series (n = 124) and a censored subset (n = 107) in which acquisitions with swath edge gaps have been removed. The low rank components of both are effectively 2D with the annual and diurnal land-water contrast accounting for ~99% of variance and vegetation-substrate contrast only ~1%. While the RPCA does relegate the swath gaps to the sparse component, some faint residual does remain in the low rank component. The residual incoherent variance introduced by the swath gaps is apparent in the difference in the higher order (>2) dimensions of the low rank components, but accounts for < 1% of total variance. In contrast, the effect of the swath gaps is most apparent in dimensions 2 and 3 of the sparse component variance – accounting for ~7% of total variance. The remaining incoherent variance related to clouds, sensor artifacts and transient thermal anomalies is distributed more evenly over the PC dimensions of the sparse component time series, with a maximum of 20% in PC1 and 14 dimensions >1% accounting for ~79% total variance. Statistically, the RPCA has very effectively separated the spatiotemporally coherent diurnal and annual thermal cycles of the land cover and water bodies from the spatiotemporally incoherent variance associated with clouds, swath gaps, sensor artifacts, transient anomalies and geospatial misregistration.

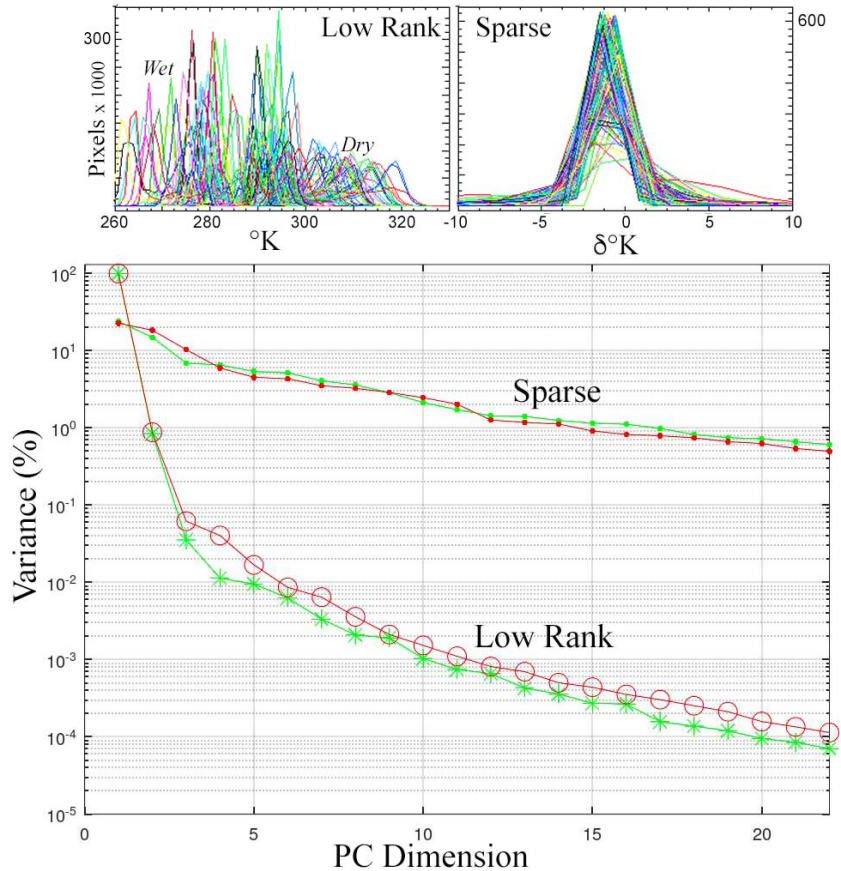

*Figure 5 Temperature distributions and variance partition for the low rank and sparse components of NYC ECOSTRESS LST time series. Low rank distributions (top left) are all bimodal while sparse residuals are generally ± 5°K. Variance partition of both full (o) and censored (*) time series show the low rank component as 2D with ~99% and 1% of variance in the first two dimensions and a continuum with < 0.1% in all higher dimensions. The residual variance of the swath gaps and geographic displacements in the full time series is apparent in the difference between the higher order dimensions. In contrast, the variance partition of the sparse residual component is more uniformly dispersed over all dimensions for both time series.*

The spatial PCs of the low rank component form a temporal feature space within which distinct temporal patterns can be identified. The temporal feature space of the three low order PCs is rendered as orthogonal scatterplots in Figure 6. The phase and amplitude differences between annual cycles of land and water are apparent in the PC1-PC2 projection and in the temporal endmembers plotted below. Water bodies are characterized by a ~30°K annual cycle peaking sharply in mid-July with near linear increase and decrease. In contrast, land surfaces show a more sinusoidal cycle peaking in early June and declining more gradually. The broad envelope of the land surface cycles reflects the much greater diurnal temperature range resulting from the lower thermal inertia of most land cover. This is apparent in the temporal feature space of PC2 and PC3. Airports (e.g. EWR, JFK) and industrial areas show the highest amplitude diurnal thermal cycles due to the large open areas that are unshaded during the day and strongly radiative at night. Forested areas have lower amplitude diurnal cycles due to the thermal inertia of leaf water content and the cooling effect of transpiration during the day. The PC3 dimension of the temporal feature space also shows the effect of a prominent NE-SW thermal gradient that seems to be driven by the different temperatures of the Long Island Sound and Lower Bay of the Atlantic Ocean respectively.

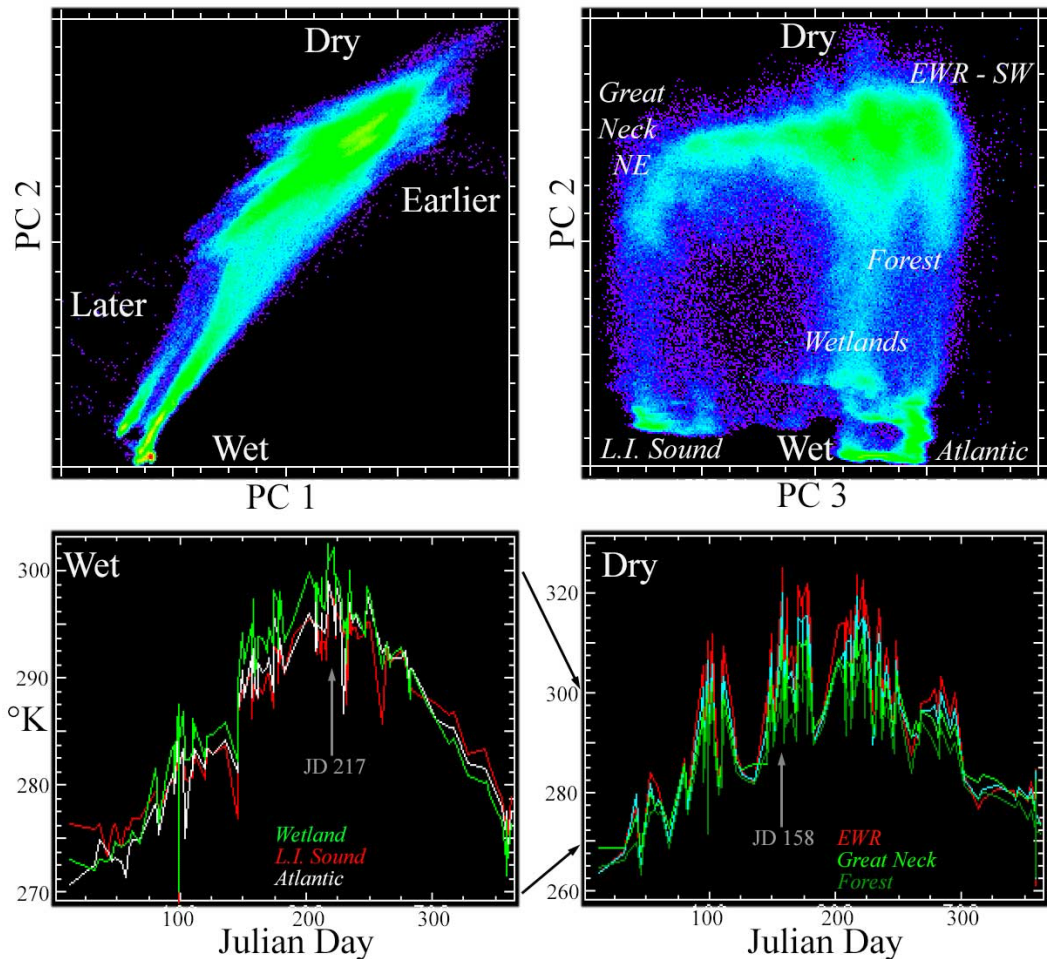

*Figure 6 Low rank component temporal feature space and temporal endmembers. Two low order dimensions of the spatial principal components illustrate the contrast between temperature range and seasonal cycle phase, while PC 3 shows NE-SW thermal gradient. Water bodies (lower left) have reduced seasonal and diurnal range with annual peaks occurring around jd217 while land areas have a greater temperature range with earlier and more prolonged seasonal peaks. Vegetated areas such as wetlands, forests and large cemeteries span the thermal gradient between wet and dry due to evapotranspiration and thermal inertia of leaf water content. The NE - SW thermal gradient in PC 3 affects both land and water bodies. Note also much greater annual and diurnal temperature range of land compared to water.*

All of the thermal cycles seen in the temporal feature space are apparent in geographic maps of the low order PCs. Figure 7 shows PC composites for the full (n = 124) low rank component time series and the subset (n = 12) of nighttime (0-10 UTC) acquisitions. As expected, the land-water contrast is prominent on both composites, with water having very low values in PCs 1 and 2. On land, the strongest contrast is between the thermal cycles of open areas like airports (red) and residential neighborhoods with high densities of deciduous tree canopy (green).  In addition to airports and industrial areas, thermal cycle hotspots are also apparent in midtown Manhattan, the Financial District (southernmost Manhattan) and downtown Brooklyn with their high concentration of tall buildings.  The most prominent small thermal hotspot is located SW of Newark Airport (EWR) adjacent to the NW corner of Staten Island.  This is an industrial area with multiple refineries and chemical plants.  As would be expected, the Port of Elizabeth on the westernmost edge of NY Harbor also stands out as a thermal hotspot.

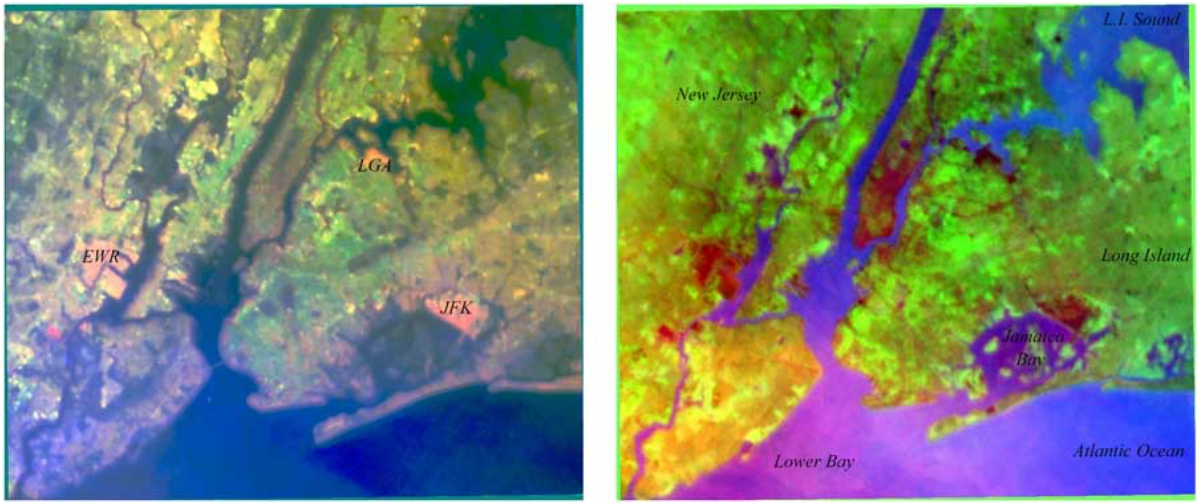

*Figure 7  Low order spatial PCs of ECOSTRESS LST low rank time series.  PCs 1 (r), 2 (g), 3 (b) of the full composite year time series (left) illustrate contrasting thermal responses of the most (red) and least (blue/black) variable surfaces.  Areas in green are more seasonally variable than darker areas.  PCs 3 (r), 1 (g), 2 (b) of the nighttime only time series (right) show a similar pattern overall, but with more pronounced nighttime hot spots in some parts of Manhattan and downtown Brooklyn.  Warmer water temperatures are also apparent for the Lower Bay, Rockaway Inlet and Verrazzano Narrows.  Gaussian stretches applied to both images to emphasize smaller thermal gradient features on land.*

*Examples of Low Rank and Sparse Component Separation*

Figures 8 – 12 provide illustrations of the RPCA separation of low rank and sparse component time series for a variety of transient thermal anomalies, sensor artifacts, swath gaps and spatial misregistrations. In each case, the raw input images are shown along with their low rank and sparse components side by side. A common 70°K temperature range (260° to 330°) is used for all raw and low rank examples with a smaller 20°K (-10° to 10°) range for the sparse residuals.

Clouds appear dark, generally saturated black in the sparse residual, in all LST acquisitions where they occur. In many cases, thin translucent clouds are apparent in the sparse residual image but are barely visible in the raw image. In no case is any cloud "ghost residual" apparent in the low rank component image, indicating that the RPCA is completely effective for cloud removal.

A variety of transient thermal anomalies are apparent in many of the sparse residual images. Most prominent are larger water body anomalies. Warm residuals are frequently observed in the shallow waters of Jamaica Bay (near JFK airport), the New Jersey Meadowlands (extensive marshes) and the shallow bays between the necks on the north shore of Long Island. Steep thermal gradients are also frequently observed between the Atlantic Ocean and the outflow plumes from New York Harbor and Jamaica Bay. Smaller, high amplitude thermal anomalies are observed throughout the metro area. Primarily associated with the aforementioned industrial areas where large, unshaded warehouse and factory roofs and staging yards are common.

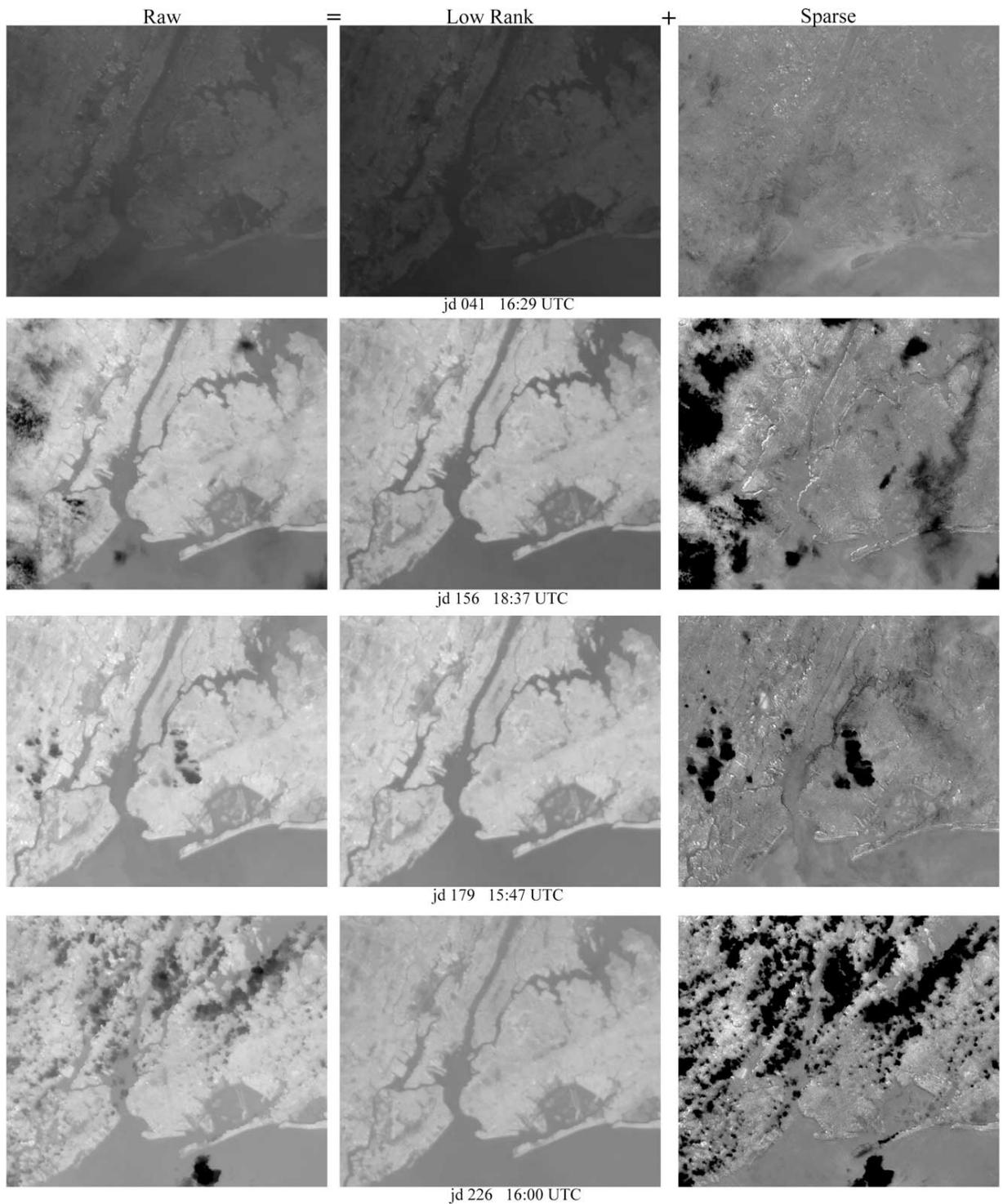

*Figure 8  Raw LST, low rank and sparse components for daytime acquisitions.  Note complete suppression of even thin, almost invisible, clouds (darker) to sparse residual.  Note also subtle anomalies in sea surface temperature plumes and small isolated hot spots.  Temperature scale is 260°K to 330°K for raw and low rank but -10°K to 10°K for residuals.*

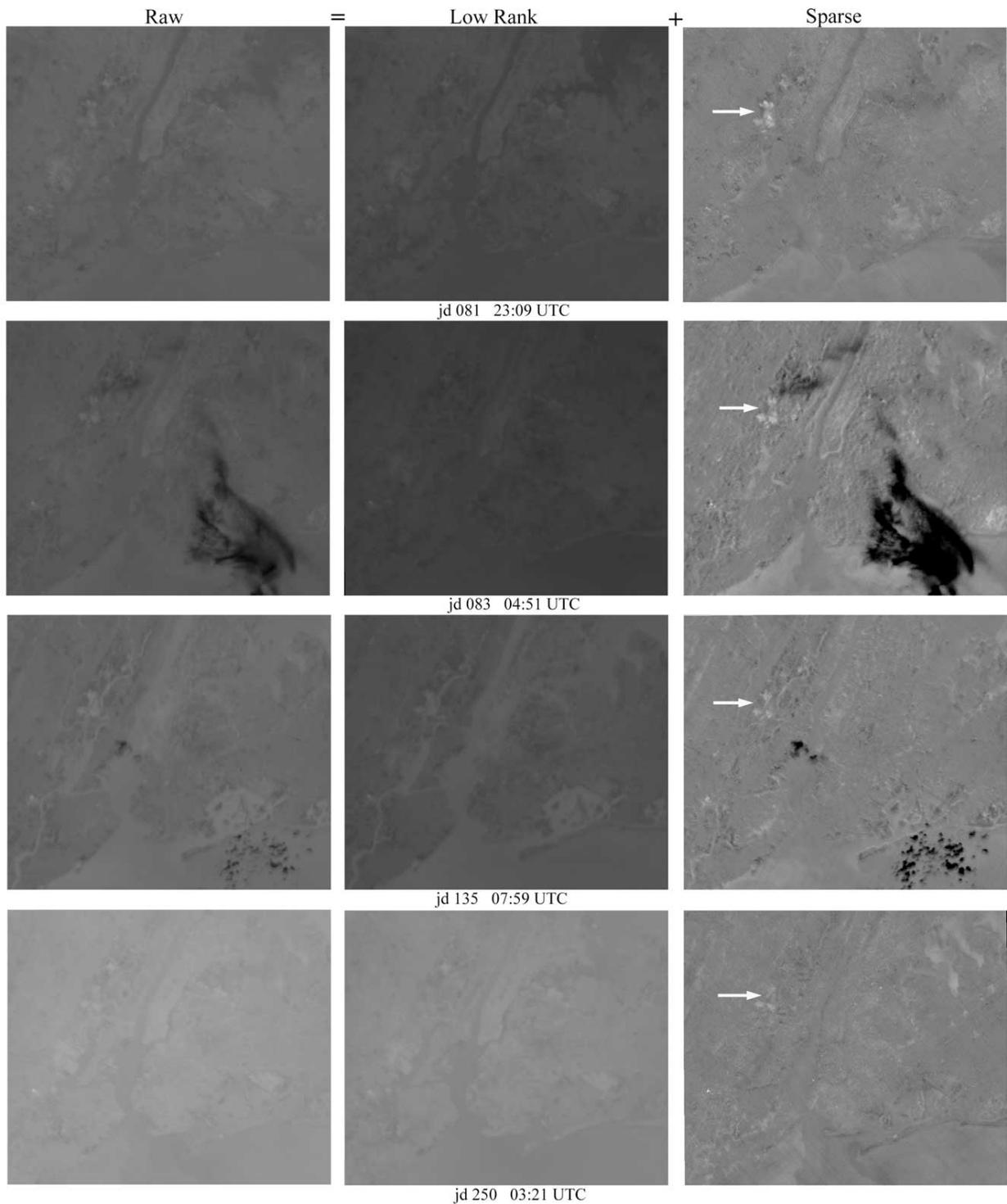

*Figure 9 Raw LST, low rank and sparse components for nighttime acquisitions. Note much lower thermal contrast between land and water compared to daytime acquisitions in Fig. 8. Diurnal thermal cycle is illustrated by reduced land-water contrast between early evening on jd081 and near midnight on jd083. Note persistent warm anomaly in shallow marsh waters of NJ Meadowlands (arrows). Temperature scales same as Fig. 8.*

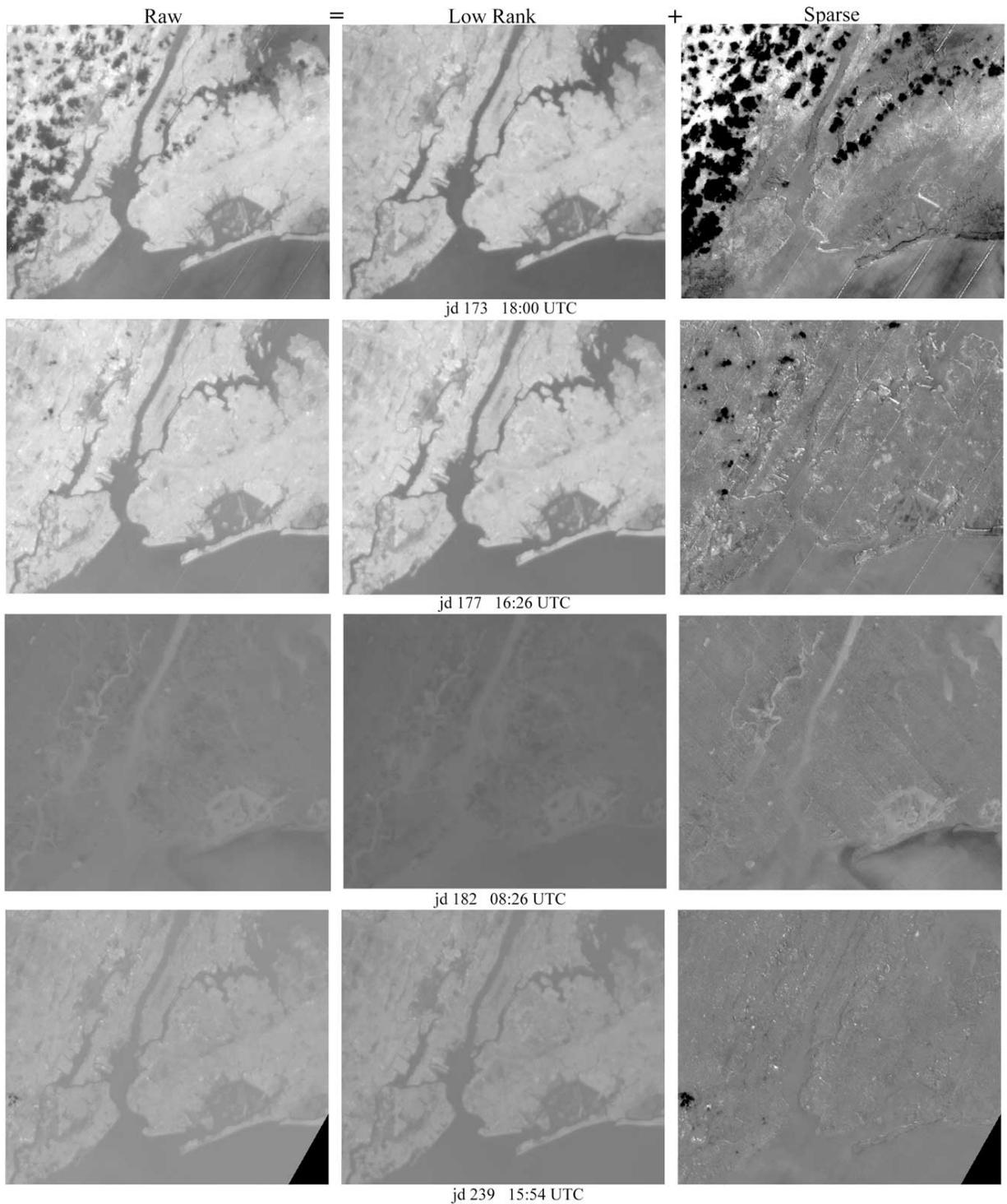

*Figure 10  Raw LST, low rank and sparse components for acquisitions containing anomalies and sensor artifacts.  The swath gap on jd239 and diagonal artifacts on the other 3 acquisitions are completely removed from the low rank components.  Numerous point source hotspots and water body anomalies are apparent in the sparse residuals.  Temperature scales same as Figs. 8 & 9.*

The effectiveness of the RPCA to compensate for swath edge gaps varies with the size of the gap.  Figure 11 shows some residual ghost darkening in the low rank component image for acquisitions with larger gaps (e.g. jd165 & jd218) where linear discontinuities can be seen. However, the compensation for smaller swath gaps appears to be effective, with no ghost effect apparent and thermal structure retained in gapped areas of the low rank images.

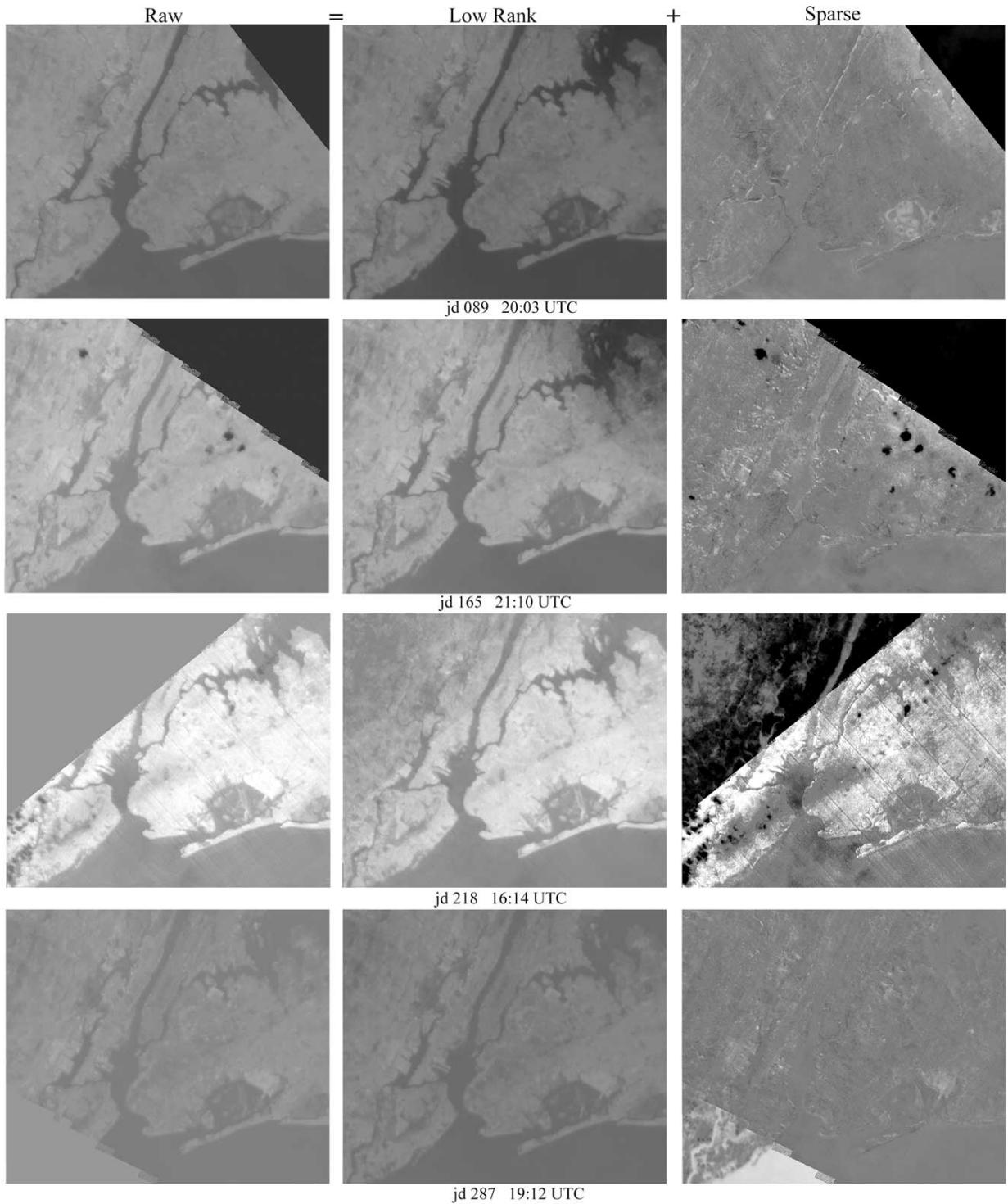

*Figure 11 Raw LST, low rank and sparse components for acquisitions containing swath gaps of different size, location, and contrast. While a distinct discontinuity is apparent on the low rank*

*components of jd 165 and jd 218, the NE corner of jd 089 and jd 165 is conspicuously darker on both low rank images. Numerous point source hotspots and water body anomalies are also apparent in the sparse residuals. Temperature scales same as Figs. 8-10.*

The RPCA is also effective for identification of spatially misregistered acquisitions. Figure 12 shows examples in which misregistered images are apparent in the complementary residual anomalies along coastlines. Because significant displacement would result in misalignment of all thermal features in the low rank component, a significant number of misregistered images would be expected to corrupt the spatiotemporal structure of the low rank time series. Therefore, images identified as misregistered in the sparse component could be either corrected (if possible) or excluded from subsequent spatiotemporal analysis. In this study, a significant number of displaced acquisitions were identified by the inability of AROSICS to achieve suitable alignment and were excluded from the outset. However, the few with significant displacements that remain in the time series were retained to illustrate the effectiveness of the RPCA to separate the displacement effects from the low rank structure of the thermal cycles.

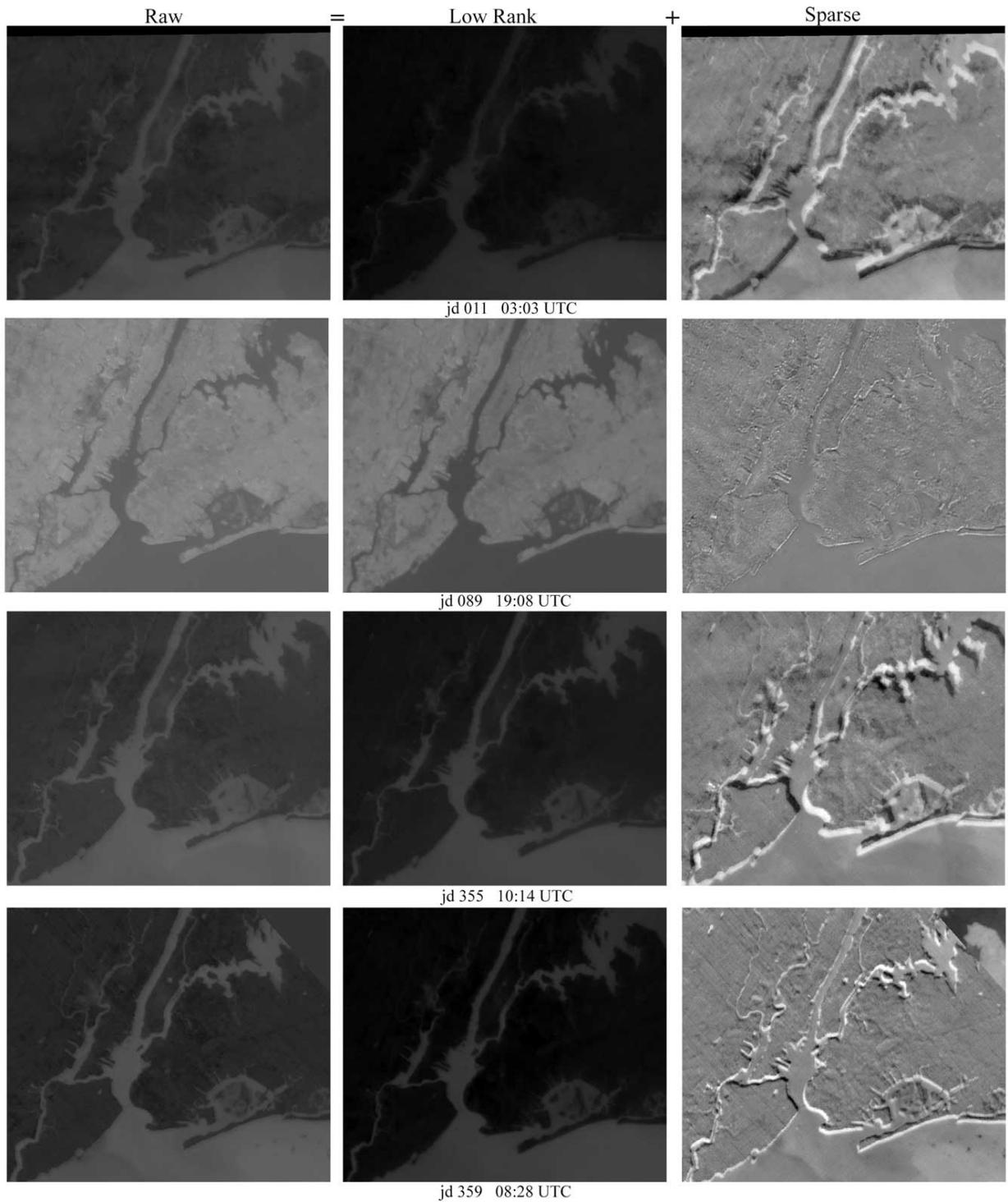

*Figure 12 Raw LST, low rank and sparse components for acquisitions with uncorrected geographic displacements. Positive and negative sparse residual anomalies along shorelines indicate direction of displacement relative to base image. Note the lack of thermal anomalies smaller than displacement anomalies in each sparse residual. Temperature scales same as Figs. 8-11.*

**Discussion**

*Process Segregation - Why It Works*

The primary spatiotemporal thermal field of the NYC study area is determined by a relatively small number of physical processes. By far, the dominant process is the differing thermal responses of land and water to both diurnal and annual cycles of solar forcing. This direct forcing, in the form of absorption of sunlight and subsequent longwave emission, is supplemented by the passage of warm and cool air masses and by advection of warm and cool water masses in the Hudson River, Atlantic Ocean and Long Island Sound. To a much smaller degree, the thermal responses of different types of land cover influences their absorption of sunlight during the day and emission of longwave radiation at night. Superimposed on these processes is the formation and movement of clouds through the study area and a variety of localized transient thermal anomalies. The measurement of all of these processes is subject to the characteristics of the imaging system, such as sensor artifacts and swath+orbital geometry giving rise to swath edge gaps.

The variance partition of the raw LST time series, given by the eigenvalues of its Singular Value Decomposition, explains less than 50% of total variance in the primary dimension of its temporal feature space, and requires more than 40 dimensions to account for 99% of total variance. In contrast, the low rank component time series requires only 2 dimensions to represent 99% of total variance, and these 2 dimensions clearly depict the physical processes driving the diurnal and annual cycles of the thermal field. By segregating the correlated low rank structure from the uncorrelated sparse transients of the spatiotemporal thermal field, the RPCA effectively isolates the predictable physical processes driving the diurnal and annual periodicities of different land covers and water masses from the more stochastic processes giving rise to transient cloud cover and sensor anomalies. In addition, the segregation of the predictable thermal cycles from the non-periodic transients makes smaller transient thermal anomalies much easier to identify. The segregation given by the RPCA distinguishes the persistent and predictable components of the thermal field that are present in consistent spatial patterns of every LST image from the transient and spatially stochastic components related to cloud cover and sensor artifacts. In this case, the RPCA works because the dominant spatiotemporal patterns present in every image can be represented with a small number of distinct spatial patterns with distinct diurnal and annual cycles, while the transient anomalies do not form coherent patterns in space or time.

*Limitations*

The factors that make RPCA so effective for spatiotemporal process segregation also give rise to its primary limitation. Like the traditional $L^2$ PCA, the effectiveness of RPCA is data-dependent. Specifically, it relies on the presence of some consistent low rank structure with sufficient variance to allow it to be distinguished statistically from the sparse structure of spatiotemporal transients. Had we included all LST acquisitions available, without regard to fractional cloud cover or geospatial rectification, the relative contribution of the low rank structure of the diurnal and annual thermal cycles would have been diminished accordingly by the spatiotemporally incoherent variance associated with higher fractions of cloud cover and reduced spatial coherence of the land and water bodies.

The asymmetric day-night distribution of usable LST acquisitions in the study area results in a significant temporal aliasing of the diurnal thermal cycle. However, the spatial similarity of diurnal and annual heating and cooling cycles combined with the strong polarity of the land and water bodies in the study area does capture enough of the diurnal cycle structure to allow it to be represented accurately in the low rank component. The three low order spatial PCs of the 12 usable nighttime acquisitions retain sufficient spatial detail to distinguish land cover types with different nocturnal cooling characteristics. This suggests that the combination of RPCA and low rank spatiotemporal characterization proposed here could potentially complement the multi-sensor diurnal+seasonal mapping of regional thermal processes such as [26-30].

*Additional Spatiotemporal Applications*

The ability of RPCA to separate pervasive annual cycles suggests that it may be useful for studies of vegetation phenology. Indeed, the combination of RPCA + spatiotemporal characterization + temporal mixture modeling has been used to suppress combined effects of cloud cover and agricultural phenologies in the study of mangrove phenology and disturbance response in the Bangladesh Sundarban [31]. Despite the subtle phenology of evergreen mangrove species, and the diversity of double cropping practices in the agricultural landscape surrounding the Sundarban, the low rank component of the RPCA was able to resolve the varying effect of pre and post monsoon phenology among different tree communities within the mangrove. However, as in this study, the near complete cloud cover Landsat and Sentinel 2 acquisitions during the monsoon were omitted from the vegetation fraction time series. Given the consistent performance of RPCA component separation with these very different types of spatiotemporal process, we expect that this approach could find application in other types of phenology analysis for which partial cloud cover is a problem as well as ground-based time-lapse thermography analysis in which transient anomalies are superimposed on a diurnal or annual cycle background. However, it is important to note that the relatively low spatiotemporal dimensionality of the thermal cycles resulting from the strongly contrasting heat capacities of land surfaces and water bodies does not necessarily apply to vegetation phenologies, which are subject to climatic variations in temperature and precipitation, as well as anthropogenic factors (e.g. agriculture). Spatiotemporal dimensionality of vegetation phenology is often on the order of 7 to 9 [13] with temporal feature spaces significantly more complex than that observed in this study [32].

**Acknowledgments**

D.S. gratefully acknowledges funding from the USDA NIFA Sustainable Agroecosystems program (Grant #2022-67019-36397), the USDA AFRI Rapid Response to Extreme Weather Events Across Food and Agricultural Systems program (Grant #2023-68016-40683), the NASA Land-Cover/Land Use Change program (Grant #NNH21ZDA001N-LCLUC), the NASA Remote Sensing of Water Quality program (Grant #80NSSC22K0907), the NASA Applications-Oriented Augmentations for Research and Analysis Program (Grant #80NSSC23K1460), the NASA Commercial Smallsat Data Analysis Program (Grant #80NSSC24K0052), the NASA FireSense airborne science program (Grant # 80NSSC24K0145), the California Climate Action Seed Award Program, and the NSF Signals in the Soil program (Award

#2226649). C.S. gratefully acknowledges the support of the endowment of the Lamont Doherty Earth Observatory of Columbia University.